\documentclass[preprint,12pt]{aastex}

\begin{document}
\title{The Evolution of the First Core in Rotating Molecular Core}
\author{Kazuya Saigo \altaffilmark{1} and Kohji Tomisaka \altaffilmark{2}} 
\affil {National Astronomical Observatory of Japan, Mitaka, Tokyo 181-8588, Japan}
\altaffiltext{1}{e-mail: saigo@th.nao.ac.jp}
\altaffiltext{2}{School of Physical Sciences, The Graduate University for Advanced Studies (SOKENDAI); e-mail tomisaka@th.nao.ac.jp }


\begin{abstract}

The standard star formation scenario is based on spherically symmetric radiative hydrodynamics calculations. 
In this paper, we focus on the evolution of the rotating non-spherical cloud through the first core and second collapse phases $10^{-13} \, < \, \rho \, < \, 10^{^3}$ g cm$^{-3}$ using axisymmetric numerical calculations. 
The rotating first core evolves under the influence of mass and angular momentum accretion from the infalling envelope. The structure of the rotating first core is characterized by the angular momentum $J_{\rm core}$ and mass $M_{\rm core}$, both of which increase with time $\dot{J}_{\rm core}$, $\dot{M}_{\rm core} \, > \, 0$. 
The evolution path of a rotating first core is considered as a sequence of equilibrium solutions with a constant  $J_{\rm core}/M_{\rm core}^2$. 
Such evolution paths are in good agreement with the results of three-dimensional numerical simulations. 
Only the slowly rotating first core with $J_{\rm core}/M_{\rm core}^2 \, < \, 0.015 G/(\sqrt{2} c_{\rm iso})$ begins the second collapse directly. 
On the other hand, the cloud with a rotation rate of $J_{\rm core}/M_{\rm core}^2 \, > \, 0.015G/(\sqrt{2} c_{\rm iso})$ does not increase in the central density beyond a certain maximum value. Although the first core grows in mass by accretion, this first core does not begin the second collapse directly. 
Such a first core has a very long time-scale and seems to suffer from a non-axisymmetric evolution, such as formation of massive spiral arms, deformation into a bar, or fragmentation.  
In the usual case, the time-scale of a rotating first core becomes several thousand years or more. So we expect that at least several percent of the prestellar cores contain a first core. The rotating first core has a small average luminosity of $L_{\rm core} \, = \, 0.003-0.03$($\dot{M}_{\rm core}/\dot{M}_{\rm ref}$) L$_{\odot}$. 
However, we expect that the rotating first core repeats temporal increases in the luminosity during its first core phase.

\end{abstract}

\keywords{theory --- hydrodynamics --- methods: numerical --- accretion disks --- stars: formation}


\section{INTRODUCTION}
\label{intro}

It is known that stars are formed through the gravitational collapse of the molecular cloud core with a density rise of about fifteen orders of magnitude. 
A dense pre-stellar core has a central density of $n \sim 10^{5}$ cm$^{-3}$  \citep{Motte98, Kandori05} and shows statistically significant inward motions \citep[see, e.g.,][]{Zhou93, Lee01}. 
It is believed that the prestellar core forms a protostar which is observed as a class 0 source in a  time-scale of $\sim$ several times of $10^{5}$ yr  by statistical studies \citep[][]{Onishi98}. 

However, their large optical depth, lack of tracer, and short time-scale prevent us from observing objects between the pre-stellar core ($n \sim  10^{5}$ cm$^{-3}$ ) and the protostar (class 0 object) phases. 
The protostar comes to be observed as a class 0 source which had already accreted significant mass from the envelope by that time. 
This missing link is one of the most important targets of next-generation observational instruments. 
Theoretically, it is necessary to make clear the process of stellar core formation from prestellar core. 
This gives the initial condition of the main accretion phase afterwards. 

The gstandardh star formation scenario from a prestellar core to a  protostar is based on the numerical simulations of spherical collapsing clouds core.  
\citet{Larson69} and \citet{Appenzeller74} calculated the spherical collapse from gas cloud to protostar by using the diffusion (or Eddington) approximations and/or the gray approximation of radiation transfer. 
\citet{Masunaga98} followed the evolution using the radiative hydrodynamical simulation and obtained similar results. 
The collapsing cloud is initially optically thin to the thermal emission from dust grains, and the cloud experiences isothermal runaway collapse. 
During this phase, the cloud collapses in a self-similar manner \citep[][]{Penston69, Larson69}. 
After that, the isothermal condition is broken at the density of $n_{\rm H2} \simeq 10^{10}$ cm$^{-3}$ and the gas temperature begins to increase.  
The collapsing cloud forms a hydrostatic equilibrium object with a radius of a few AU and mass of $\sim 0.01 M_{\odot}$, which is called the gfirst core.h 
The first core increases its mass by accretion from the isothermal envelope and it raises the central density and temperature. 
The hydrogen molecules begin to dissociate into atoms after the central temperature reaches about 2000 K ($n_{\rm H2} \simeq 10^{16}$ cm$^{-3}$). 
Since this reaction extracts the thermal energy from gas, a gsecond collapseh begins. 
After the dissociation is completed ($n_{\rm H2} \sim 10^{20}$ cm$^{-3}$), a gsecond coreh(or gstellar coreh) is born at the center.

However, interstellar cloud cores show rotation motions.   
\citet{Goodman93} observed 43 cloud cores by NH$_{3}$ $(J, K) = (1, 1)$ transition  ($n_{\rm H2} \simeq 10^{4}$ cm$^{-3}$) and found that the angular velocity $\Omega_{\rm c}$ of these cores is distributed in the range of $(0.3 - 4) \times 10^{-6}$ yr$^{-1}$. 
\citet{Caselli02} obtained similar results of $\Omega_{\rm c} \, = \, (0.5 - 6) \times 10^{-6}$ yr$^{-1}$ with N$_{2}$H$^{+}$ data ($n_{\rm H2} \simeq 10^{5}$ cm$^{-3}$). 
These rotations correspond to the ratio of the rotational energy to the self-gravitational energy of $T/|W| \sim 10^{-4} - 0.07$. 
Rotation of the cloud core must have a crucial influence on star formation \citep[for a review see][]{Bodenheimer00}. 

Even today, computational power is not sufficient for multi-dimensional radiative hydrodynamics, and therefore a number of simplifications are supposed in such investigations. 
Isothermal gas is a good approximation for the gas of $n_{\rm H2} < 10^{10}$ cm$^{-3}$ as well as the spherical case. 
Cloud cores near hydrostatic equilibrium are hard to fragment during the isothermal runaway collapse phase \citep[][]{Boss87,Boss95,Miyama92,Tsuribe99}. 
During this phase, the central region of the collapsing cloud spins up and it forms a dynamically collapsing disk \citep[][]{Norman80,Miyama84}. 
\citet{Matsumoto97} have shown that such a collapsing isothermal cloud tends to have a central angular velocity of  $\Omega_{\rm c} \sim (0.2 - 0.3)/ t_{\rm ff}$ in the runaway collapse phase, where $t_{\rm ff}$ is the free-fall time scale at the central density. 

After the isothermal condition is broken, many authors follow the subsequent evolution by assuming the polytropic equation of state based on the results of spherical cloud collapse \citep{Bonnell94a, Bate98} or radiative transfer in diffusion approximation \citep{Boss86, Boss95}. 
\citet[][hereafter MH03]{Matsumoto03} followed the evolution using the nested grid  method with a polytropic equation of state. 
The first core accretes mass and angular momentum from the infalling envelope and thus the radius and mass of the first core are much larger than the non-rotating spherical model.   The radius and mass of the rotating first core are equal to a few tens AU and a few times 0.01 $M_{\odot}$, respectively.
They found that the first core also increases in $T/|W|$ by mass accretion. 
The polytropic gas disks with $T/|W| \, > \,  0.27$ suffer from axisymmetric instability \citep[][]{Durisen86, Pickett96, Imamura00} under a wide range of conditions. 
\citet{Bonnell94a} confirmed that the first core with $T/|W| \, > \, 0.3 $ becomes unstable against the non-axisymmetric instability to both $m \, = \, 1$ and 2 modes during the accretion phase and forms massive arms. 
Subsequently, the first core increases in its density by angular momentum transport via gravitational torque or fragmentation into binary or multiples. 
MH03 did an extensive parameter survey of the collapse of rotating Bonner-Ebert clouds with a central density of $n_{\rm H2} \, = \, 2.6 \times 10^{4}$ cm$^{-3}$. 
They found that first cores have a wide variety of morphologies depending on the initial rotation speed, $\Omega_{\rm c} t_{\rm ff}$ and that clouds with $\Omega_{\rm c0} \, \ge \, 0.05 / t_{\rm ff}$ fragment into binaries or multiples in the first core phase.  

Thus, the accretion of mass and angular momentum leads to a crucial influence on the evolution of the first core and first cores must have a wide variety of evolution paths depending on the angular momentum of the infalling matter. 
However, it is not clear how the evolution path of the first core is systematically affected by mass accretion. 
In this paper, we show the evolution of the first core by using equilibrium solutions  comprehensively and quantitatively. 

In a non-rotating case, the first core begins to collapse again, and form a stellar core. 
This second collapse is important in understanding star formation, because it presents the direct initial condition of the protostar and protostellar disk and it may lead to formation of a close binary. 
However, a lot of obscurities and discussions remain concerning the evolution of a rotating cloud. 
\citet{Boss89} and \citet{Bonnell94b} calculated the second collapse phase from a gravitationally unstable rotating first core, assuming a polytropic equation of state. 
They found that the fragmentation occurs through two mechanisms once the collapse  is halted. 
They found two typical evolutions: 
first, a protostellar disk fragments by non-axisymmetric $m  \, = \, 1$ and 2 modes. 
Second, a collapsing cloud bounces centrifugally and forms a ring at the end of the second collapse phase. This ring results in fragments of several pieces. 
However, they assumed a first core with uniform density and Gaussian density profile instead of a rotational equilibrium configuration.   This must affect the evolution. 

On the other hand, \citet{Bate98} obtained a different result. 
He followed the cloud collapse from the isothermal phase to the second core formation by the SPH method, assuming a polytropic equation of state. 
In his calculation, the first core forms two massive arms during the first core phase and a large amount of angular momentum is removed from the core. This results in a slowly rotating core and forms a stellar core directly, without any fragmentations, similar to the non-rotating spherical collapse. 
However, the number of models in his letter is so restricted, it is difficult to understand how widely these results are realized. 
We examine hydrodynamical evolution of the second collapse from the end of the first core phase, which is  obtained by our quasi-static calculation. From extensive parameter study of rotation parameters we show the evolution is classified into three types according to rotation and density.

The phases of this paper are as follows: the model of the rotating first core is described in \S 2, and we show the quasi-static evolutional path of the first core with mass accretion from the envelope in \S 3. In \S 4, we explore the second collapse more comprehensively with hydrodynamical simulations. 
We discusses the results in \S 5.

\section{ROTATING FIRST CORE}
\label{sec2}

\subsection{Basic Equations and Numerical Method}

We calculate the equilibrium state of the rotating first core with various central densities and rotation speeds in order to obtain its evolution path  as a series of solutions. The basic equations are as follows, 
\begin{equation}
(\rho r \Omega^2, 0, 0) - \nabla P - \rho \nabla \psi=0,
\label{mod-forc}
\end{equation}
and 
\begin{equation}
\Delta \psi = 4 \pi G \rho
\label{mod-pois}
\end{equation}
where \mbox{\boldmath$v$}, $p$, $\psi$, and $G$ denote 
the velocity, thermal pressure, gravitational potential, and gravitational constant, respectively. 
All the variables are assumed to be functions of $r$ and $z$, 
and independent of $ \varphi $ in the cylindrical coordinates $(r, \varphi, z)$.

For simplicity, we did not solve the energy equation which must include radiative cooling  and chemical reaction. 
Instead, we assume a polytropic equation of state, 
\begin{equation} 
P \, = \, K \rho^{\gamma} \; ,
\label{mod-poly}
\end{equation}
where $\gamma$ and $K$ are the polytropic index and constant, respectively.  To mimic the thermal history of the first core, we switched $\gamma$ and $K$ at the $H_{2}$ dissociation density, $\rho_{\rm diss} \, = \, 5.6 \times 10^{-8}$ g cm$^{-3}$,  as follows,
\begin{equation}
(\gamma, K) \,  = \, \left\{
\begin{array}{ccll} 
(7/5, \, 5.697 \times 10^{13}) \; &  & \rho 
< &\rho_{\rm diss}  , \\
(1.1, \, 3.823 \times 10^{11}) \; & \rho_{\rm diss} <& \rho 
&, 
\end{array}
\right.
\label{mod-gamm}
\end{equation}
in cgs unit.
The temperature is proportional to $T \, \propto \, \rho^{\gamma - 1}$ in this polytropic model. 
The transition density and polytropic parameters are taken from the radiative hydrodynamical calculations of \citet{Masunaga98}.

\subsection{Models}

The rotating first core is expressed as an axisymmetric equilibrium of a rotating polytropic gas cloud with a mass of $M_{\rm core}$ and  total angular momentum of $J_{\rm core}$.
We assumed that the distribution of angular momentum of the rotating first core is the same as one for a uniformly rotating, uniform-density sphere,
\begin{equation}
j(M(R))\, = \, \frac{5}{2}\left(\frac{J_{\rm core}}{M_{\rm core}}\right) \left\{ 1 - \left(1-\frac{M(R)}{M_{\rm core}}\right)^{2/3} \right\},
\end{equation}
where $j(M(R))$ is the local specific angular momentum at cylindrical radius $R$ from the rotational axis, $M(R)$ is the mass within the cylindrical radius of $R$.

Density and rotation velocity distribution of the first core are numerically  constructed by the self-consistent field (SCF) method of \citet{Hachisu86}, which iterates between solutions of the Poisson equation and that of the equilibrium equation among pressure, centrifugal, and gravitational forces. 
We extend the original SCF code to include the effect of polytropic index $\gamma$ dependent on density (see eq. [\ref{mod-gamm}]). 
The density is assumed to equal to zero outside the first core for simplicity. 

The equilibria of rotating first cores are specified by only two model parameters, the central density, $\rho_{\rm c0}$, and the total angular momentum, $J_{\rm core}$, according to \citet{Hachisu86}. 
The mass of the first core $M_{\rm core}$ changes depending on these two model parameters. 
Figure \ref{fig1} illustrates the density distributions of the first cores with the same central density as $\rho_{\rm c0} \, = \, 4 \rho_{\rm diss}$ but three different angular momenta. They are typical initial conditions of evolution models. 
Their total angular momenta are $J_{\rm core} \, = \, 9.99 \times 10^{49}$, $4.18 \times 10^{49}$, and $2.25 \times 10^{49}$ g cm$^{2}$ s$^{-1}$ and hereafter we call them Models A, B, and C, respectively. 
The contour curves and the gray scale show the logarithm of the density.  
White dashed curves denote the iso-density contour at $\rho \, = \, \rho_{\rm diss}$. Figure \ref{fig1} shows that all first cores form density plateaus in the central part and have concave-shaped  envelopes.  The first core is larger and more massive for models with larger angular momentum. 
Their masses are $M_{\rm core} \, = \,  4.97 \times 10^{31}$,  $3.45 \times 10^{31}$ and $2.77 \times 10^{31}$ g. 
The thicknesses of their cores are identical 0.4 AU because the cores are supported only by the pressure force in the $z$-direction.

\section{Evolution Path of Rotating First Core with Mass Accretion}
\label{contraction}

During the first core contraction phase, the mass and the central density of the rotating first core increases gradually by means of mass accretion from the infalling isothermal envelope.

Figure \ref{fig2} illustrates the relation between the central density, $\rho_{\rm c0}$, and the mass of the first core, $M_{\rm core}$. 
The thick curve denotes a series of non-rotating solutions with $J_{\rm core} \, = \, 0$. The non-rotating first core evolves along this curve from left to right as the mass increases by accretion from the infalling envelope. When it reaches the maximum mass of $\simeq$ 0.01 $M_{\odot}$ at $\rho_{\rm c0} \, \simeq \, 2  \rho_{\rm diss}$, further mass accretion results in a dynamical collapse, since there is no hydrostatic equilibrium beyond this mass. The cloud goes into the second collapse phase \citep[see][]{Larson69}. 
This comes from the dependency of the Jeans mass, $M_{\rm J} \, \propto \, T^{3/2} \rho^{-1/2} \, \propto \, \rho^{(3 \gamma-4)/2}$. 
The mass of the first core increases in proportion to $\rho_{\rm c0}^{0.1}$ in the low density part $\rho_{\rm c0} \, \la \, 10^{-7}$ g cm$^{-3}$ and decreases in proportion to $\rho_{\rm c0}^{-0.35}$ in the high density part $\rho_{\rm c0} \, \ga \, 10^{-7}$ g cm$^{-3}$.

In the rotating case, the first core becomes more massive by support of the centrifugal force against gravitational force. 
The evolution path is not only controlled by mass accretion but also angular momentum accretion from the infalling envelope. Thus, there are innumerable evolution paths which have different increases in angular momentum $J_{\rm core}$. 
Here, we assume the mass accreting first core evolves while maintainin  the value of $J_{\rm core}/M_{\rm core}^{2}$. 
It is because many numerical simulations show that the specific angular momentum in the central region of the rotating cloud tends to be proportional to the enclosed mass during the isothermal collapse \citep{Matsumoto97}. 
We introduce a non-dimensional rotation rate, $\omega$, as an index of the rotation  defined as 
\begin{equation}
\omega \, \equiv \, \frac{\sqrt{2} c_{\rm iso}}{G} \, \frac{J_{\rm core}}{M_{\rm core}^2}, 
\end{equation} 
where $c_{\rm iso}$ is the isothermal sound speed. 
Each solid curve in Figure \ref{fig2} denotes the series of the equilibrium solution with the same $\omega$ and denotes the evolution paths of the first core with the infalling isothermal envelope.

Figure \ref{fig2} indicates that the central density increases with the core mass for $\rho_{\rm c0} \, \la \, \rho_{\rm diss}$. 
In the case of $\omega \, < \, 0.015$, the rotating first core has a maximum mass at $\rho_{\rm c0} \, = \, (2 - 4)  \rho_{\rm diss}$ as well as the non-rotating case. 
These is no hydrostatic equilibrium if the mass is larger than this maximum value. Such a first core begins the dynamical collapse, which is called the second collapse after the mass exceeds the maximum mass. 
The solutions with $\partial M / \partial \rho_{\rm c0}|_{\omega} \, < \, 0$ are unstable, and are indicated with a region colored in red in Figure \ref{fig2}.

In contrast, the mass of the first core increases monotonically with the central density for $\omega \, > \, 0.015$ and the unstable region $\partial M / \partial \rho_{\rm c}|_{\omega} \, < \, 0$ disappears.   
Thus, the first core cannot begin the second collapse, even if $M_{\rm core} \, > \, 0.018 M_{\odot}$ at $\rho_{\rm c0} \, \simeq 4 \rho_{\rm diss}$. 
This means that the maximum mass of the core that experiences the second collapse is 0.018 $M_{\odot}$. 
The central density hardly increases even if the accretion increases the mass. 
Such a first core dose not begin the second collapse unless the core spinning slows down owing to the fragmentation or the angular momentum transport.


\section{SECOND COLLAPSE PHASE}

\subsection{Numerical Method and Boundary Conditions}

We obtained the quasi-static evolutionary path of the first core caused by the mass and angular momentum accretion. 
The end points of the quasi-static evolution have been shown in $\S 3$.
Now, we calculate the hydrodynamical evolution of the rotating first core by numerical simulations. 
The initial conditions are the rotating first core obtained in \S 2 and \S 3.
We examine the stability of the first core and follow the gravitational collapse during the second collapse phase. 
We superpose 1\% density enhancement on the cloud and assume a low-density gas of  $\sim 10^{-5} \rho_{\rm diss}$ outside the first core.  

The basic equations for hydrodynamical calculation are as follows, 
\begin{equation}
\frac{\partial \rho}{\partial t} + \nabla \cdot 
(\rho \mbox{\boldmath$v$}) = 0 , 
\label{hydro-mcon}
\end{equation}
\begin{equation}
\frac{\partial}{\partial t}(\rho \mbox{\boldmath$v$}) + 
\nabla \cdot (\rho \mbox{\boldmath$v$} \otimes \mbox{\boldmath$v$})
 + \nabla P + \rho \nabla \psi=0,
\label{hydro-forc}
\end{equation}
and 
\begin{equation}
\Delta \psi = 4 \pi G \rho
\label{hydro-pois}.
\end{equation}
All the variables are assumed to be functions of $r$ and $z$, 
and independent of $ \varphi $ 
in the cylindrical coordinates $(r, \varphi, z)$.
For simplicity, we did not solve the energy equation and we assume a polytropic relation similar to equation  (\ref{mod-poly}) as  
\begin{equation}
\gamma \, = \, \left\{
\begin{array}{ccll} 
1.0 \; && \rho < &\rho_{\rm iso} \, = \, 10^{-13} \, {\rm g} \,{\rm cm}^{-3},     \\
7/5 \; &  \rho_{\rm iso} <& \rho 
< &\rho_{\rm diss} \, = \, 5.6 \times 10^{-8} \, {\rm g} \,{\rm cm}^{-3},\\
1.1 \; & \rho_{\rm diss} <& \rho 
< &\rho_{\rm stel} \, = \,  10^{-3} \, {\rm g} \, {\rm cm}^{-3},\\
5/3 \; & \rho_{\rm stel} <& \rho  &,
\end{array}
\right.
\label{hdro-gamm}
\end{equation}
where $\rho_{\rm iso}$ and $\rho_{\rm stel}$ are the transition density from the isothermal to adiabatic equation of states and the density after which the hydrogen molecule dissociation is completed.

The calculations were performed using a two-dimensional axisymmetric Euler code. We developed the numerical code to solve the hydrodynamical equations (\ref{hydro-mcon})-(\ref{hydro-pois}) by modifying that of \citet{Saigo00}. 
Our code consists of two parts, the solver of the hydrodynamical equations and that of the Poisson equation. The former is based on the Flux Difference Splitting (FDS) scheme of \citet{Roe81}. The second order accuracy is achieved by the Monotone Upwind Scheme for Conservation Law (MUSCL) method \citep[e.g.][]{Hirsch90}.
We solved the Poisson equation with the five-point central difference scheme and the multi-grid iteration \citep[see, e.g.][]{Press91}.

We achieve high spatial resolution throughout the computations by rezoning.
As the central density increases with the gravitational collapse, the central high-density region shrinks. 
To compute the small-scale structure, we resize the grid cell and restrict the computation domain to the central part with the grid number of $(n_r, n_z)\, = \, (512, 512)$ in the $r$- and $z$-directions. 
The size of the initial computational domain is taken to be twice as large as the disk radius.

The first core is assumed to be isolated and we assume no flux across the outer boundary. The boundary condition of the gravitational potential is set by the spherical harmonics expansion. 

We introduce a new finer grid so that the resolution satisfies the condition that the half-width of the half-maximum density should be resolved with more than 16 grid spacings. 
The size of the finer computational domain is taken to be half that of the coarser one.
When a new grid is introduced, the density and velocity on the grid points are interpolated by the first order bilinear method. 
The density, velocity and gravitational potential are fixed on the outer boundary after rezoning. The Poisson equation is solved on the fine grid with the boundary gravitational potential at the rezoning epoch.

\subsection{Typical Evolutions}

Here, we show the dynamical evolution of the rotating clouds with a wide range of rotation speeds from the end of the first core to the stellar core. The evolution of the first core shown to be is divided into three types according to rotation speed. 

\subsubsection{Model A: Direct Stellar Core Formation}

In this subsection, we show the second collapse of the slowly rotating cloud of model A. 
It is a typical example in which the runaway second collapse continues till the stellar core formation. 
The parameters of model A are $\rho_{\rm c0} \, = \, 4 \rho_{\rm diss} \, = \, 2.3 \times 10^{-7}$ g cm$^{-3}$ and $\omega \, = \, 0.012$, which has the maximum allowable mass with the same rotation rate, $\omega$ (Fig. \ref{fig3}{\it a}). 
This model shows evolutions qualitatively similar to those of the non-rotating spherical collapse model \citep{Larson69, Masunaga98}.

Figure \ref{fig3}{\it b} shows the density distribution at the age of $ t \, =\, 4105.3$ days after the dynamical contraction begins. 
The first core begins runaway collapse with the gravitational instability and the central density $\rho_{\rm c}$ is ten times larger than that of the initial condition of dynamical evolution, $\rho_{\rm c0}$. In this paper, we defined the collapse in which the central density rises by one order of magnitude as the second collapse. 
In the early stage of second collapse, the first core collapses mainly vertically. Note that the dynamical collapse is seen not only in the central dissociation region  $\rho \, > \, \rho_{\rm diss}$ but also in the outer part of the first core. 

Further evolution was traced with finer grids. 
Figure \ref{fig3}{\it c} and \ref{fig3}{\it d} shows the density distributions at the ages of $t \, =\, 4519.3$ days and  $t \, =\, 4531.7$ days, respectively. 
The central densities are 100 and 1000 times larger than the initial one,  respectively. 
The gas disk evolves self-similarly during the runaway collapse phase.
The central high density disk becomes thinner and thinner during the runaway collapse. 
This second collapse phase ends and the increase in the central density stops at $t \, =\, 4546.5$ days (Fig. \ref{fig3}{\it e}). 
The white solid curve denotes the stellar core of $\rho_{\rm c} \, > \, \rho_{\rm stel}$.
The stellar core bounces outwards on the equatorial plane \citep[see][]{Bonnell94b}. 
At the stage of $t \, =\, 4601.8$ days, the density of the core has fallen below the stellar one and the core is surrounded by a shock wave with a radius of $0.02$ AU (Fig. \ref{fig3}{\it f}).

Figure \ref{fig4}{\it a} shows the density profiles on the equatorial plane. 
It is seen each density profile consists in a plateau around the center and a power-law envelope and the evolution proceeds in a self-similar fashion. 
During the runaway collapse phase (solid curves), the density in the envelope is proportional to $r^{-2.68}$. 
This power-law index is steeper than that of the spherical collapsing polytropic gas of $r^{-2/(2 - \gamma)} \, = \, r^{-2.22}$ ($\gamma \, = \, 1.1$) \citep[see e.g.][]{Suto88}. 
It is due to the increase in the ratio of the centrifugal force to the pressure gradient force during runaway collapse. 

During calculation, Toomre $Q$-value inside the central plateau of the collapsing cloud is below unity. 
We also plot the unstable wavelength of $\lambda_{\rm J}$ in figure \ref{fig4}{\it a} which is marked as a cross. 
The unstable length is derived from the Toomre reference analysis for the centrifugally supported disk, 
\begin{equation}
\lambda_{\rm J} \, = \, \frac{2 c_{\rm s}^{2}}{G \Sigma (1 + \sqrt{1-Q^{2}})}
\end{equation}
where $c_{\rm s}$, $\Sigma$, and $Q$ denotes the sound speed, surface density of the disk, and Toomre $Q$-value, respectively. 
It is equal to or less than the radius of the central high density plateau on each profile, so the collapsing disk is stable against the axisymmetric instability. 

Figures \ref{fig4}{\it b} and \ref{fig4}{\it c} show the infall and rotation velocity profiles on the equatorial plane. 
They have maxima at the edge of the central density plateau and decrease with radius in the envelope. 
This distribution is the same as the self-similar solution of the spherical collapsing polytrope \citep{Yahil83,Suto88}. 
Figure \ref{fig4}{\it d} shows the infall velocity profiles along the $z$-axis. 
The maximum infall velocity attains $2.5 \times 10^{5}$ cm s$^{-1}$ at the edge of the central high density region.

\subsubsection{Model B: Formation of Rotationally Supported Core}

Even when the rotation is a little faster than model A, the evolution becomes quite different from model A. We show model B as a typical example for runaway collapse models with larger angular momenta. 
The initial condition of model B (Fig. \ref{fig5}{\it a}) is $\rho_{\rm c0} \, = \, 4 \rho_{\rm diss}$ and $M_{\rm core} \, = \, 3.45 \times 10^{31}$ g ($\omega \, = \, 0.014$). 

Figures \ref{fig5}{\it b} and \ref{fig5}{\it c} show the density distribution at $ t \, =\, 4982.9$ and 5712.9 days. The central density of these stages are 10 and 100 times higher than the initial state. 
As in model A, the cloud begins its runaway collapse as a whole with the gravitational instability. 

In this model, the second collapse suddenly stops and the cloud forms  an equilibrium core (or disk) before the second core formation. Figure \ref{fig5}{\it d} shows the disk (inside the white dotted curve) at the stage of $\rho_{\rm c} \, = \, 2.45 \times 10^{-5}$ g cm$^{-3}$ which is much lower than the stellar density of $\rho_{\rm stel} \, = \, 1.0 
\times 10^{-3}$ g cm$^{-3}$ (see eq. [\ref{hdro-gamm}]). 
Here, we define the equilibrium core as the area where the infall speed becomes less than 20\% of sound speed, which may be regarded as the 1.5th core.
The core is mainly supported by centrifugal force. The force fraction of the centrifugal to the pressure force is 0.65. 
The radius and mass of the 1.5th core is 0.02 AU and $8.8 \times 10^{30}$ g, respectively. 
At the stage of Figure \ref{fig5}{\it e}, the radius and mass of the disk increase to 0.18 AU and $1.05 \times 10^{31}$ g due to mass accretion, respectively. 
On the other hand, the density of the central region decreases due to the centrifugal bounce.

Figure \ref{fig6}{\it a} shows the density profiles on the equatorial plane, $\rho (r)$. 
This shows that the evolution proceeds roughly self-similarly as well as model A. During the runaway collapse phase (solid curves), the density in the envelope is proportional to $r^{-2.9}$. 
This profile is steeper than that of model A. 
During the accretion phase (dashed curves), the central density decreases to about one-half of the maximum value at $t \, = \, 5734.8 $ days due to a bounce of the disk. 
The centrifugally supported disk accretes mass through a shock front  which is seen at $r \simeq 0.16$. 

Figures 6{\it b} and 6{\it c} show the radial infall, $v_{r}$, and rotation velocities, $v_{\varphi}$, on the equatorial plane.  
The evolution is almost similar to model A. 
At stage ($\it d$), the infall velocity has vanished inside the radius of 0.02 AU.
It is the radius of the high density plateau. 
Driven by the accretion, the disk radius increases to 0.16 AU at stage ({\it e}).

\subsubsection{Model C: Stable First Core}

In faster rotating model C, the rotating first core indicates an oscillation around the initial equilibrium state and does not show any indications of the runaway collapse. 
Model C is a typical example of such a stable first core.
The parameters are $\rho_{\rm c0} \, = \, 4 \rho_{\rm diss}$ and $M_{\rm core} \, = \, 4.97 \times 10^{31}$ g (see Fig. \ref{fig2}).

\subsection{Classification of Second Collapse Phase}

The initial conditions of hydrodynamical calculations are plotted with crosses, triangles, and circles in Figure 2. 
First cores marked by open crosses collapse dynamically until formation of the stellar core as shown in model A. 
These models are distributed below the curve of $\omega \, = \, 0.012$ in Figure 2. 
First cores marked by triangles collapse dynamically but form centrifugally supported disks below the stellar density. 
Rapidly rotating cores indicated by crosses do not experience the second collapse phase.
This non-collapse model is defined as the model in which the central density does not increase to 10 times that of the initial one.  
The boundary between the collapse model and non-collapse model roughly corresponds to the boundary which separates the stable $\partial M / \partial \rho_{\rm c} \, > \, 0$ and unstable $\partial M / \partial \rho_{\rm c} \, < \, 0$ regions already shown in \S 3 (red region in Fig. 2).

\section{DISCUSSION}

\subsection{$T/|W|$}

Figure 7 is the same as Figure 2 but it shows the rotational energy of the first core. 
The color scale shows the ratio of rotational energy to the gravitational energy, $T/|W|$. 
The first core with larger rotational energy $T/|W|$ can support more mass. 
A dashed curve denotes the critical curve of $T/|W| \, = \, 0.273$ for non-axisymmetric instability.
That is, \citet{Imamura00} showed that the rotating polytropic gas disk with $\gamma \, = \, 7/5$ is unstable against the non-axisymmetric instability if $T/|W| \, > \, 0.273$ \citep[see also,][]{Durisen86, Pickett96}

The first cores with $\omega \, > \, 0.015$ stop increasing central density, and the evolution path rises nearly vertically with mass accretion. 
During the evolution, $T/|W|$ increases monotonically along the path. 
Finally $T/|W|$ exceeds the critical value 0.273 and the first core enters the non-axisymmetric unstable region.  
On the other hand, the first core with $\omega \, < \, 0.015$ evolves to the right in Figure 7 and begins the second collapse directly without non-axisymmetric instability. In this case, $T/|W|$ increases slowly and there is a maximum value of 0.2 at most.

\subsection{Comparison with Three-dimensional Calculations}

A number of authors have studied the formation and evolution of the first core from the rotating cloud with various initial conditions of rotation and density distribution using three-dimensional calculations \citep[][]{Bonnell94a,Bate98,Saigo02,Matsumoto03}. 
They found that the first core shows a wide variety of evolutions depending on its initial condition. 
However, its evolution had not been understood comprehensively. 
Here, we show that the variety of results is understood using the central density – core mass ($\rho_{\rm c0}-M_{\rm core}$) diagram (Fig. \ref{fig7}). 

Numerical simulations have show that the first core which is formed from a rotating cloud becomes non-axisymmetrically unstable in a short time after the formation of the first core.  
This can be understood if we think that all the clouds in their calculations have sufficiently fast rotation of $\omega \, > \, 0.015$. 

\citet{Bate98} calculated the collapse of a uniformly rotating ($\Omega_{\rm c} \, = \, 7.6 \times 10^{-14}$ s$^{-1}$) uniform density ($\rho \, = \, 10^{-18}$ g cm$^{-3}$) sphere using the SPH method with a polytropic equation of state. 
The rotating cloud forms a hydrostatic first core with a mass of $M_{\rm core} \simeq 0.01 M_{\odot}$ at $\rho_{\rm c} \, \simeq \, 2 \times 10^{-11}$ g  cm$^{-3}$ (lower square symbol in Fig. \ref{fig7}). 
After that, a non-axisymmetric unstable $m \, = \, 2$ mode begins to grow at $\rho_{\rm c} \, \simeq \, 10^{-10}$ g cm$^{-3}$. 
At that time, the first core has $M_{\rm core} \simeq 0.04 M_{\odot}$, $R_{\rm core} \simeq 40$ AU and $T/|W| \simeq 0.34$ (upper square symbol in Fig. \ref{fig7}). 
Between these stages, the cloud evolves nearly vertically in Figure \ref{fig7}. 
The evolution almost agrees with our evolution path with $\omega \, = \, 0.06$. 
This value roughly agrees with the initial value $\omega \, \simeq \, \Omega_{\rm c} t_{\rm ff} = 0.08$ at $\rho \, = \, 10^{-18}$ g cm$^{-3}$ (See Appendix A).

Afterward, this first core forms two massive spiral arms and the angular momentum is transferred to the outer disk (see Figs. 3 and 5 of Bate 1998). The central region begins the second collapse and it continues till the stellar core formation. 
In our analysis, the first core with $\omega \, = \, 0.06$ can start the second collapse if $J_{\rm core}/M_{\rm core^{2}}$ of the core becomes 1/4 or less of the value before the formation of spiral arms. \citet{Bate98} reported that the total angular momentum of the central disk within several AU decreases to $\sim$ 20\% of the value before the formation of spiral arms \footnote{The exact change of the rotation rate $\omega$ angular has not been obtained from \citet{Bate98} owing to a lack of information about the mass within several AU.}.

MH03 calculated the formation and evolution of the first core from the rotating Bonner-Ebert sphere for 225 models with different initial rotation speeds, rotation laws, and amplitudes of density perturbation. 
They found that a cloud core with non-dimensional angular speed of $\Omega_{\rm c} t_{\rm ff} \, < \, 0.03$ at the central density of $\rho_{\rm c} \, = \,  1.0 \times 10^{-19}$g cm$^{-3}$ does not fragment in their calculation. 
These first cores deform into spirals or a bar due to the non-axisymmetric instability just after the first core formation.  

In Figure 7, we plot the evolution path of their High-Resolution-Model with $\Omega_{\rm c} t_{\rm ff} \, = \, 0.0276$ initial at the central density of $\rho_{\rm c} \, = \,  1.0 \times 10^{-19}$g cm$^{-3}$ (hereafter we refer it as `HRM003') by the dashed curve (MH03). 
The central density increases gradually due to mass accretion for $\rho \le 10^{-9}$ g cm$^{-3}$. 
The evolution of HRM003 is almost along our evolution path of $\omega \, = \, 0.04$. 
After the evolution path goes over our critical curves for non-axisymmetric instability, massive spiral arms appears in the first core (Fig. 3{\it d} of MH03). 
Soon after, the central region of the spiral arms shrinks into a compact centrifugally supported core and the central density increases by an  order of magnitude (Fig. 3{\it e} of MH03). 
Thus, their result of HRM003 agrees well with our evolution path of $\omega \, = \, 0.04$ and it also agrees with our stability analysis  for non-axisymmetric instability. 
After that, the central density does not increase securely and keeps its values  $\rho_{\rm c} \, \simeq 10^{-8}$ g cm$^{-3}$, although it increases temporally by non-axisymmetric deformation of the first core. 
Note that the mass of the first core increases steadily by mass accretion. 
It seems that such non-axisymmetric instabilities in the first core lead to the  re-distribution of the angular momentum to support larger masses [we will discuss this separately in Saigo, Matsumoto, \& Tomisaka (2005)]. 
We expect that this massive first core fragments finally when it reaches the true maximum (mass) beyond which no hydrostatic configuration exists. 
A large-scale three-dimensional calculation must be performed to understand the evolution in later stages.

On the other hand, the rotating cloud core with $\Omega_{\rm c} t_{\rm ff} > 0.03$ at the central density of $\rho_{\rm c} \, = \, 1.0 \times 10^{-19}$ g cm$^{-3}$ fragments promptly in the early stage of the first core phase (MH03). 
Here, we refer to their High-Resolution-Model with $\Omega_{\rm c} t_{\rm ff} \, = \, 0.046$ at the central density of $\rho_{\rm c} \, = \,  1.0 \times 10^{-19}$g cm$^{-3}$ as a typical fragmentation model (hereafter we refer to it as `HRM005').
This rapidly rotating cloud forms the first core with a mass of $M_{\rm core} \simeq 0.02 M_{\odot}$ at $\rho_{\rm c} \, \simeq \, 6 \times 10^{-11}$ g cm$^{-3}$ (Fig.7{\it b} of MH03). 
We plot it in our Figure 7 with a filled circle symbol. 
The position is in a non-axisymmetric unstable region on the evolution path with $\omega \, \simeq \, 0.06$. 
Soon after, this first core deforms into a bar and fragments into a binary (Fig.7{\it d} of MH03). 
These fragments are self-gravitating and these high-density clumps are embedded in the extended  first core. 
To follow and understand the subsequent evolution of the fragments in our Figure 7, we define the mass of the clump, $M_{\rm clump}$, integrating the gas with $\rho > 3.8 \times 10^{-11}$ g cm$^{-3}$.
The bar fragments into two clumps with masses of $M_{\rm clump} \, = \, 1.0 \times 10^{-2} M_{\odot}$ and $1.1 \times 10^{-2} M_{\odot}$ at $\rho_{\rm c} \, \sim \, 4 \times 10^{-10}$ g  cm$^{-3}$. Each fragment accretes mass and grows $M_{\rm clump} \, = \, 3.0 \times 10^{-2} M_{\odot}$ and $2.4 \times 10^{-2} M_{\odot}$ at $\rho_{\rm c} \, \sim \, 4 \times 10^{-8}$ g  cm$^{-3}$ (Fig.7{\it f} of MH03). 
We plot these fragments in Figure 7 with open circles. 
This evolution can be understood as follows: in the fragmentation process the specific angular momentum for spin motion from each fragment is much reduced compared with the orbital angular momentum and as a result, the central density of each clump rises rapidly. 
After a while, each fragment enters the non-axisymmetric unstable region again by mass accretion. 
It is consistent with the appearance of spiral arms in each fragment shown in Figure.7{\it f} of MH03. 

It seems that through the fragmentation process a significant part of the angular momentum of the first core goes into the orbit angular momentum and the spin angular momentum of fragments is reduced. 
To estimate the angular momentum redistribution in the fragmentation process, we introduce a model core with power-law distribution of the surface density, $\Sigma ( r )$ and rotation velocity $v_{\varphi} (r)$ as 
\begin{equation}
\begin{array} {cc}
\Sigma ( r ) \, \propto \, r^{\alpha},  \\
v_{\varphi} ( r ) \, \propto \, r^{\beta}, 
\end{array} 
\end{equation}
where $r$ denotes the cylindrical radius. 
And we assume that the gas in the disk in an opening angle of $\phi_{\rm f}$ forms a fragment without any tidal interaction. 
The ratio of the spin-specific angular momentum of fragment, $ j_{\rm f, spin} / j_{\rm core}$, becomes 
\begin{equation}
\frac{j_{\rm f, spin} }{ j_{\rm core}} \, = \, 1 \, - \, \left[\frac{2 \sin (\phi_{\rm f}/2)}{\phi_{\rm f}}\right]^2 \frac{(2 + \alpha)(3 + \alpha + \beta)} {(3 + \alpha)(2 + \alpha + \beta)}
\end{equation}
(see Appendix B). 
Here, we consider the first core with the uniformly rotating and uniform surface density ($\alpha \, = \, 0$, and $\beta \, = \, 1$) as a model core. 
If the whole disk fragments into equal mass binary ($\phi _{\rm f} \, = \, \pi$), the ratio $j_{\rm f, spin} / j_{\rm core}$ becomes 0.64. The ratio decreases to 0.125 for $\phi_{\rm f} \, \ll \, 1$. 
Thus the spin-specific angular momentum, $j$, decreases to 10\% - 60\% of the specific angular momentum of the first core before fragmentation $j_{\rm f, spin}$. 
This value is consistent with the results of three-dimensional simulation.  
For example, Machida et al. (2005) obtained the ratio of $j_{\rm f, spin} / j_{\rm core} \, = \, 0.1 - 0.5$ from their three-dimensional simulations of the collapse of six magnetized clouds (the fragment is defined as high density clumps with $n \, > \, 2.5 \times 10^{11}$ cm$^{-3}$). 

Finally, we discuss the possibility that the fragmentation induces the second collapse. 
We take the fragmentation model of MH03 shown before as the initial first core (circle symbol in Fig. 7). 
From equation (6), $\omega$ of the fragments changes in proportion to the change in specific angular momentum, $j_{\rm f, spin} / j_{\rm core}$ and in inverse proportion to the fraction of mass contained in $\phi_{\rm f}$ . 
Therefore, after the fragmentation, each self-gravitating fragment  forms a new rotating first core with rotation speed of $(\phi_{\rm f}/2 \pi)^2 (j_{\rm f, spin} / j_{\rm core}) \omega$. 
We consider the lowest limit of $\omega$ as the value of model RHM003 which is a no-fragmentation model of MH03. 
If it fragments into an equal mass binary, if $j_{\rm f, spin} \, < \, 0.05 \, j_{\rm core}$, the fragments begin the second collapse. 
This redistribution fraction is smaller than the usual value of $j_{\rm f, spin} \, = \, ( 0.1 - 0.5 ) \, j_{\rm core}$. 

As a result, although the fragmentation can reduce the spin angular momentum of the cores, it is hard to induce the second collapse in the early stage of the first core.

\subsection{Effect of Magnetic Breaking}

Basu \& Mouschovias (1994) pointed out that magnetic braking plays an important role in the case of the magnetized cloud. 
Tomisaka (1998, 2000, 2002) followed the collapse of rotating magnetized cloud cores with axisymmetric magnetohydrodynamical (MHD) simulations using a two-dimensional nested grid method. 
At the end of the isothermal runaway collapse, the central region within $\simeq$ 0.01$M_{\odot}$ has a rotation rate of $ \omega \, \simeq \, 0.071$ (Fig.1 of Tomisaka 2000). 
During the first core accretion phase, the azimuthal magnetic tension transfers its angular momentum from the disk mid-plane to the surface and gas with excess angular momentum near the surface being ejected by the centrifugal force. 
The outflow is accelerated by the magnetic tension and pressure. 
This shows that the magnetized rotating first core decreases its angular momentum and continues to contract till the second collapse begins as well as the non-rotating spherical cloud. 
Tomisaka (2002) mentioned that about 99\% of the angular momentum is removed by the outflow with 10\% in mass during the first core phase and the first core begins the second collapse. 
Thus, the magnetic braking moves the equilibrium position on the $\rho_{\rm c0}-M_{\rm core}$ plane to the right. 

It should be noted that his calculations assumed an ideal MHD, although the coupling between the magnetic field and the gas weakens over the density $\rho \, > \, 4 \times 10^{-12}$ g cm$^{-3}$ (Nakano \& Umebayashi 1986). 
Because the ionization fraction of high-density gas is quite low, the electric conductivity decreases as collapse proceeds. 
In the late phase of the first core, the magnetic braking may become ineffective. 
Therefore, the ideal MHD calculation may overestimate the effect of the magnetic field. 
On the other hand, non-magnetic hydrodynamical simulations give the opposite limit.

\subsection{New Evolution Stages Induced by Rotation}

\subsubsection{Another Evolutionary Path to Form the 2nd Core without Second Collapse Phase}

To begin the second collapse, it is necessary that in the first core a significant amount of angular momentum is extracted at once from the central region. 
Here, we propose that there are other possibilities for formation of a stellar core without the second runaway collapse phase. 
There are two kinds of evolution. 
First, the central density reaches the stellar density while the first core continually loses angular momentum due to fragmentation or gravitational torque. 
Second, the gas with small angular momentum accretes on the first core in the later stage and the first core contracts gradually until the stellar core is formed.


\subsubsection{The Centrifugally Supported Disk}

In our calculation, the dynamically collapsing first core forms a rotationally supported 1.5th core before formation of the second core in models with $0.012 <  \omega < 0.015 $ at $\rho_{\rm c} \, = \, 4 \rho_{\rm diss}$. 
\citet{Bonnell94b} found the centrifugal bounce of the second core at the end of the second collapse phase in the models far from virial equilibrium initially (small  rotational support). 
This leads to a self-gravitating ring with $\rho \, < \, \rho_{\rm stel}$. 
The centrifugally supported 1.5th core may lead to a binary formation and may change the mass accretion rate onto the second core (or star) in the early phase. 


\subsection{Dependence of Angular Momentum Distribution}

We considered the evolution of the first core by assuming the angular momentum 
distribution of equation (5).
Here, we evaluate the effect of the angular momentum distribution.
Figure 8 shows the $\rho_{\rm c0}-M_{\rm core}$ diagram for the cloud with angular momentum of a uniformly rotating, uniform-density cylinder, 
\begin{equation}
j(M(R))\, = \, \left(\frac{J_{\rm core}}{M_{\rm core}}\right) \frac{M(R)}{M_{\rm core}}. 
\end{equation}

Comparing Figure 7, there are no essential differences in evolution path, gravitational stability, and the distribution of $T/|W|$. 
Therefore, the mass-central density relation is insensitive to the angular momentum distribution.

\subsection{The Luminosity  Evolution of Rotating First Core}

The evolution of luminosity and the time-scale of the first core are important quantities for the observation. Figure \ref{fig9} shows the evolutionary curves of luminosity for models of $\omega \, = \, 0.0, 0.012, 0.015, 0.016, 0.02, 0.03$, and $0.04$. The luminosity of first core $L_{\rm core}$ is obtained by an energy change as  
\begin{equation}
L_{\rm core} \, = \, - \frac{d}{dt}[W + E_{\rm Kin} + E_{\rm T}],
\end{equation}
where $W$, $E_{\rm Kin}$ and $E_{\rm T}$ denote the gravitational energy, kinematical energy and thermal energy of the first core, respectively. 
Here, we assume the constant mass accretion rate of $\dot{M} \, = \, 1.0 \times 10^{-5}$ M$_{\odot}$/yr as a reference. 
It is the typical mass accretion rate of a runaway collapsing rotating disk \citep[][]{Saigo98}. 
The time $t$ from the formation epoch is $t \, = \, M_{\rm core}/\dot{M}$. Therefore, the $t$ and $L_{\rm core}$ can be scaled by the mass accretion rate, $t \, \propto \, \frac{1}{\dot{M}_{\rm core}}$ and $L_{\rm core} \, \propto \dot{M}_{\rm core}$.
In the non-rotating case of $\omega \, = \, 0$, the time-scale of the first core contraction phase is about 1,000($\dot{M}_{\rm core}/\dot{M}_{\rm ref}$)$^{-1}$yr. The luminosity of the first core increases up to 0.03 ($\dot{M}_{\rm core}/\dot{M}_{\rm ref}$) L$_{\odot}$ at the end of the first core phase.
This is consistent with \citep[][]{Masunaga98}.

In the case of a rotating first core, the time-scale becomes much longer. 
However, its luminosity becomes smaller for larger $\omega$, because the central density becomes smaller for larger $\omega$ and the gravitational potential becomes shallower for larger $\omega$. The luminosity of the rotating model of $\omega \, = \, 0.03$ is $L_{\rm core} \, \simeq 0.01$ ($\dot{M}_{\rm core}/\dot{M}_{\rm ref}$) L$_{\odot}$. 
On the other hand, the luminosity curves for models with $\omega \, = \, 0.012$ and 0.015 show the rapid and temporal increase at $t \, \simeq \, 1500-2000$ yr in Figure \ref{fig9}. This corresponds to the rapid contraction around $\rho_{\rm c0} \simeq \rho_{\rm 4 diss}$ (see the horizontal path of fig. \ref{fig7}). 

The usual first core has a rotation with $\omega \, > \, 0.015$. So actually the time-scale of the first core is several thousand years or more depending on the accretion rate. The time-scale of the prestellar core of $n_{\rm H2} \, \sim \, 10^{5}$ cm$^{-3}$ is $\sim 0.1$M~yr \citep[][]{Onishi98}. Therefore, we expect that at least several percent of prestellar cores contain first cores and this percentage is inversely proportional to the accretion rate. 
During such a long first core phase, they must suffer from  non-axisymmetric instability and this causes rapid contraction of first cores (see evolution path of MRN003 in fig. \ref{fig7}). 
Therefore, we expect that first cores repeat temporal increases in  luminosity during their long first core phase, though it has a small average luminosity.

\section{SUMMARY}

\label{summary}

We investigated the evolution of the rotating cloud during the first core and the second collapse phases. The evolution path of first core was constructed as a sequence of the equilibrium solutions. 
The evolution is characterized by the rotation rate $\omega$. 
A cloud with an extremely slow rotation of $\omega \, < \, 0.015$ evolves similarly to the spherical collapse during the first core phase. 
On the other hand, a cloud with $\omega \, > \, 0.015$ has a certain maximum density and the first core does not begin the second collapse directly. 
The first core grows more and more in mass by mass accretion and gradually becomes   non-axisymmetrically unstable. 
Our evolution path agrees well with the results of three-dimensional calculations. 
After that, the evolution of the first core is characterized by non-axisymmetric evolution, such as a formation of massive spiral arms, deformation into a bar, or fragmentation.  
Our $\rho_{\rm c0}-M_{\rm core}$ diagram helps to explain the evolution comprehensively. 
We expect that the first core with rotation does not begin the second collapse phase without loss of a significant amount of angular momentum. 
Therefore, there is another possibility that the first core contraction phase continues gradually till formation of the stellar core.  
During the second collapse phase, a cloud with $0.015 \, < \, \omega < \, 0.012$ forms a centrifugally supported disk before the stellar core formation. 

We expect that the rotating first core has a the time-scale of several thousand years or more and at least several percent of prestellar cores contain first cores. 
The rotating first core has a small average luminosity of $L_{\rm core} \, = \, 0.003-0.03$($\dot{M}_{\rm core}/\dot{M}_{\rm ref}$) L$_{\odot}$. 
However, we expect that the rotating first core repeats temporal increases in luminosity during its first core phase.

\acknowledgments

We wish to express our gratitude to I. Hachisu who offered his numerical code (SCF) in which the equilibrium solutions were calculated. This research became possible with his code. We wish to express our gratitudes to T. Matsumoto and T. Hanawa who helped our hydrodynamical calculations through the offer of the code and data. 
We also thank S. Inutsuka,  R.B. Larson, Y. Li, M.-M. MacLow, J. Oishi, and K. Omukai for valuable discussion and comments. 
Numerical calculations and data analysis were carried out with Fujitsu VPP5000 at the Astronomical Data Analysis Center, National Astronomical Observatory of Japan. This research was supported in part by Grants-in-Aid for Scientific Research from MEXT, Japan (16204012 and 14540233[KT]).

\appendix

\section{RELATION BETWEEN NON-DIMENSTIONAL ANGULAR SPEED AND THE ROTATION RATE $\omega$}

The rotation rate $\omega$ can be rewritten using the central angular speed $\Omega_{\rm c}$ and central density $\rho_{\rm c}$, when the disk has the uniform rotation, uniform surface density, $\Sigma$ and isothermal hydrostatic equilibrium in the $z$-direction. 
In this case, the rotation rate is
\begin{equation}
\omega \, = \, \frac{\sqrt{2} c_{\rm iso} J_{\rm core}}{G M_{\rm core}^{2}} \, = \, \frac{ \sqrt{2} c_{\rm iso} \Omega_{\rm c}}{2 \pi G \Sigma},  
\end{equation}
where $c_{\rm iso}$ is the isothermal sound speed. 
In the isothermal hydrostatic equilibrium, the uniform surface density, $\Sigma$, is related to the central density as 
\begin{equation}
\Sigma \, = \, ( \frac{2 c_{\rm iso}^{2} \rho_{\rm c}}{ \pi G} )^{1/2}.
\end{equation}
If we define the free-fall time-scale as $t_{\rm ff} \, = \,  1/(4 \pi G \rho_{\rm c})^{1/2}$, the rotation rate is rewritten by non-dimensional angular speed,  
\begin{equation}
\omega \, = \, \Omega_{\rm c} t_{\rm ff}.
\end{equation}

\section{DISTRIBUTION OF ANGULAR MOMEMTUM TO ORBITAL AND SPIN}

We assume a simple model of the first core with power-law distributions of the surface density, $\Sigma (r)$ and rotation velocity, $v_{\varphi} (r)$ as  
\begin{equation}
\begin{array} {cc}
\Sigma ( r ) \, = \, A r^{\alpha}, \\
v_{\varphi} ( r ) \, = \, B r^{\beta},
\end{array} 
\end{equation}
where $r$ denotes the cylindrical radius. For $\alpha \, > \, -2$, mass of the disk is expressed as 
\begin{equation}
M_{\rm core} \, = \, \int_0^{R_{\rm core}} 2 \pi r \Sigma ( r ) dr 
\, = \, \frac{2 \pi A}{\alpha + 2} R_{\rm core}^{ \alpha + 2},
\end{equation}
where $R_{\rm core}$ is the outer radius of the first core. The average  specific angular momentum of the first core becomes 
\begin{equation}
j_{\rm core} \, = \, \frac{\int_0^{R_{\rm core}} 2 \pi r \Sigma ( r ) r v_{\varphi} dr}{M_{\rm core}} 
\, = \, \frac{\alpha + 2}{\alpha + \beta + 3} B R_{\rm core}^{ \beta + 1}.
\end{equation}
We assume that the disk fragments and gas in the opening angle of $\phi_{\rm f}$ forms a compact object (fragment) without any tidal interaction. The mass,  position and velocity of the fragment are  $M_{\rm f} \, = \,(\phi_{\rm f}/2\pi) M_{\rm core}$, $\mbox{\boldmath$r$}_{\rm f} \, = \, \int_{\rm core} \Sigma (\mbox{\boldmath$r$})~\mbox{\boldmath$r$} \; r dr d\phi /M_{\rm f}$, and 
$\mbox{\boldmath$v$}_{\rm f} \, = \, \int_{\rm core} \Sigma (\mbox{\boldmath$r$}) ~\mbox{\boldmath$v$} ~ r dr d\phi /M_{\rm f}$, respectively, where we take the foot of the position vector to be at the disk center.  
They can be rewritten as 
\begin{equation}
\begin{array} {lll}
r_{\rm f} \, & = &\, |\mbox{\boldmath$r$}_{\rm f}| \, = \,  \displaystyle{ M_{\rm f}^{-1}
\int_{0}^{R_{\rm core}} \int_{(\pi-\phi_{\rm f})/2}^{(\pi+\phi_{\rm f})/2} \Sigma (r) ~r \; r \sin{\phi} d\phi  dr  
             }
\, = \, \frac{2 \sin (\phi_{\rm f}/2)} {\phi_{\rm f}} \frac{\alpha + 2}{\alpha + 3} R_{\rm core}, \\ 
v_{\bot} \,& = &\, 
\displaystyle{
 M_{\rm f}^{-1} \int_{0}^{R_{\rm core}} \int_{(\pi-\phi_{\rm f})/2}^{(\pi+\phi_{\rm f})/2} \Sigma (r) ~v_{\varphi}(r)~ \; r \sin{\phi} d\phi  dr
            } \, = \,  
\frac{2 \sin (\phi_{\rm f}/2)} {\phi_{\rm f}} \frac{\alpha + 2} {\alpha + \beta + 2} R_{\rm core}^{\beta}\\
v_{\|} \, & = & \, 0
\end{array}
\end{equation}
where $v_{\bot}$ and $v_{\|}$ denote the perpendicular and parallel components of velocity to the position vector $\mbox{\boldmath$r$}$, respectively. 
The specific orbital angular momentum of the fragment is $j_{\rm f, orbit} \, = \, r_{\rm} \times v_{\bot}$. 
Therefore, the ratio of the spin-specific angular momentum of fragment, $ j_{\rm f, spin} / j_{\rm core}$, becomes 
\begin{equation}
\frac{j_{\rm f, spin} }{ j_{\rm core}} \, = \, 1 \, - \, \left[ \frac{2 \sin (\phi_{\rm f}/2)} {\phi_{\rm f}} \right]^{2} \frac{(2 + \alpha)(3 + \alpha + \beta)} {(3 + \alpha)(2 + \alpha + \beta)}.
\end{equation}




\setcounter{figure}{0}

\clearpage
\begin{figure}
\plotone{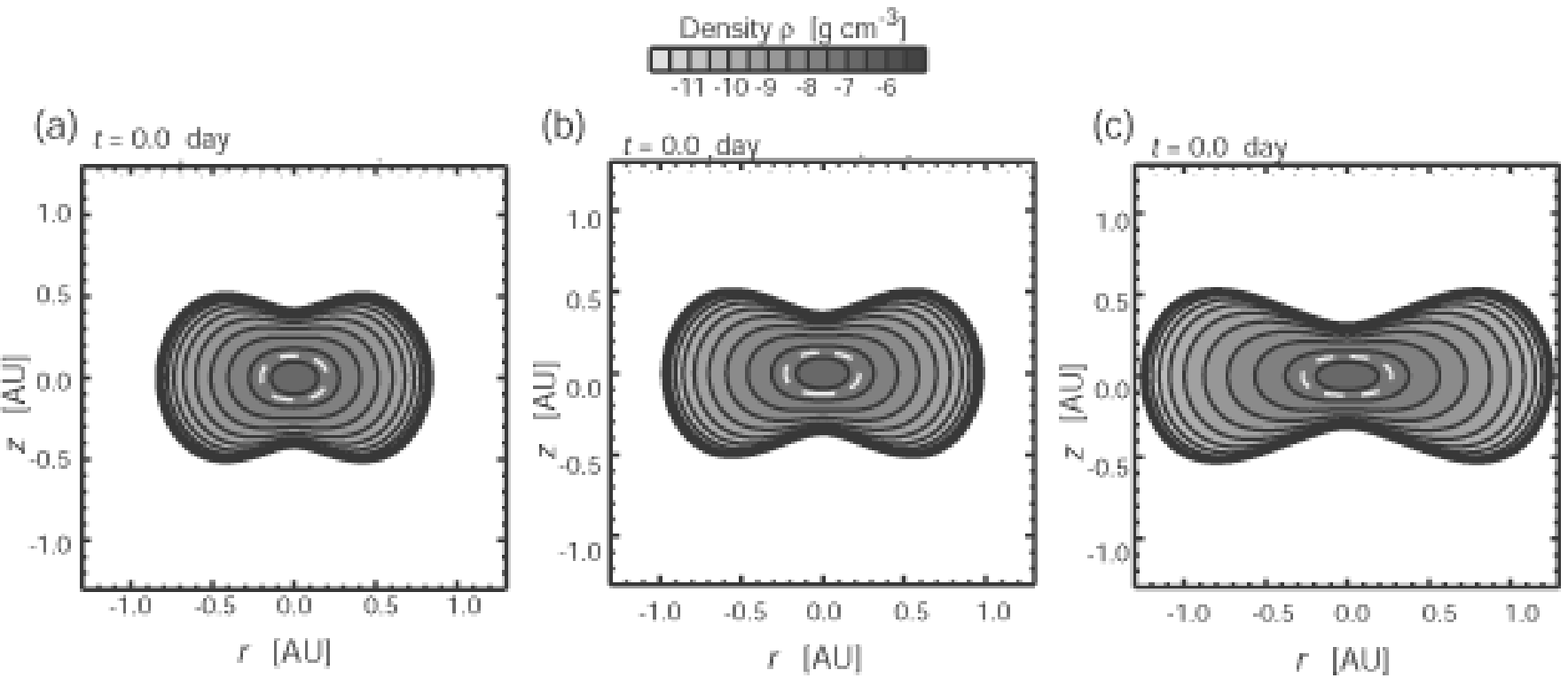}
\caption[dummy]{
Rotating first cores with the same central density of $\rho_{\rm c0} \, = \, 4 \rho_{\rm diss}$ but different rotation speeds.
The contours and gray scale denote the density distribution in logarithmic scale on the $r$-$z$ plane. The white dashed curve denotes the isodensity contour of constant density with dissociation density $\rho_{\rm diss}$. 
They are examples for ({\it a}) slow rotation with $T/|W| \, = \, 0.11$ (Model A), ({\it b}) moderate rotation with $T/|W| \, = \, 0.15$ (Model B), and  ({\it c}) fast rotation with $T/|W| \, = \, 0.22$ (Model C), where $T/|W|$ is the ratio of rotational energy to the gravitational energy. 
\label{fig1}}
\end{figure}

\begin{figure}
\plotone{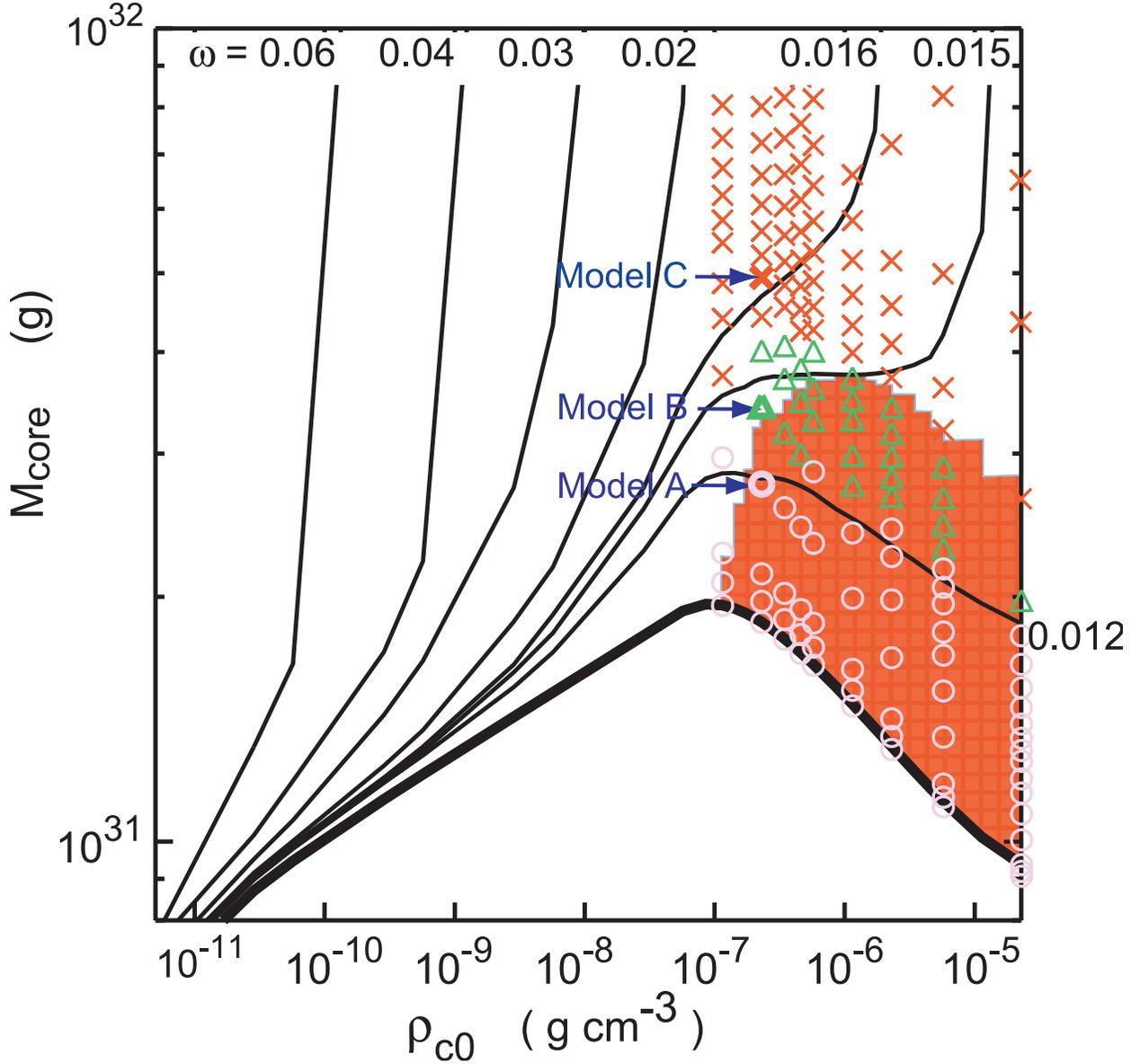}
\caption[dummy]{
The central density – core mass ($\rho_{\rm c0}$-$M_{\rm core}$) diagram. The evolution path is plotted as sequences of equilibrium solutions.
The solid curves denote the evolution path and each path is labeled by rotation rate $\omega$. The thick curve denotes the evolution path of the non-rotation spherical model. 
The red region shows the gravitationally unstable region. 
Circles denote first cores which form second cores directly.
Triangles denote first cores which form centrifugally supported disks before the formation of the stellar core. 
Crosses indicate stable models with no contraction. 
\label{fig2}}
\end{figure}

\begin{figure}
\plotone{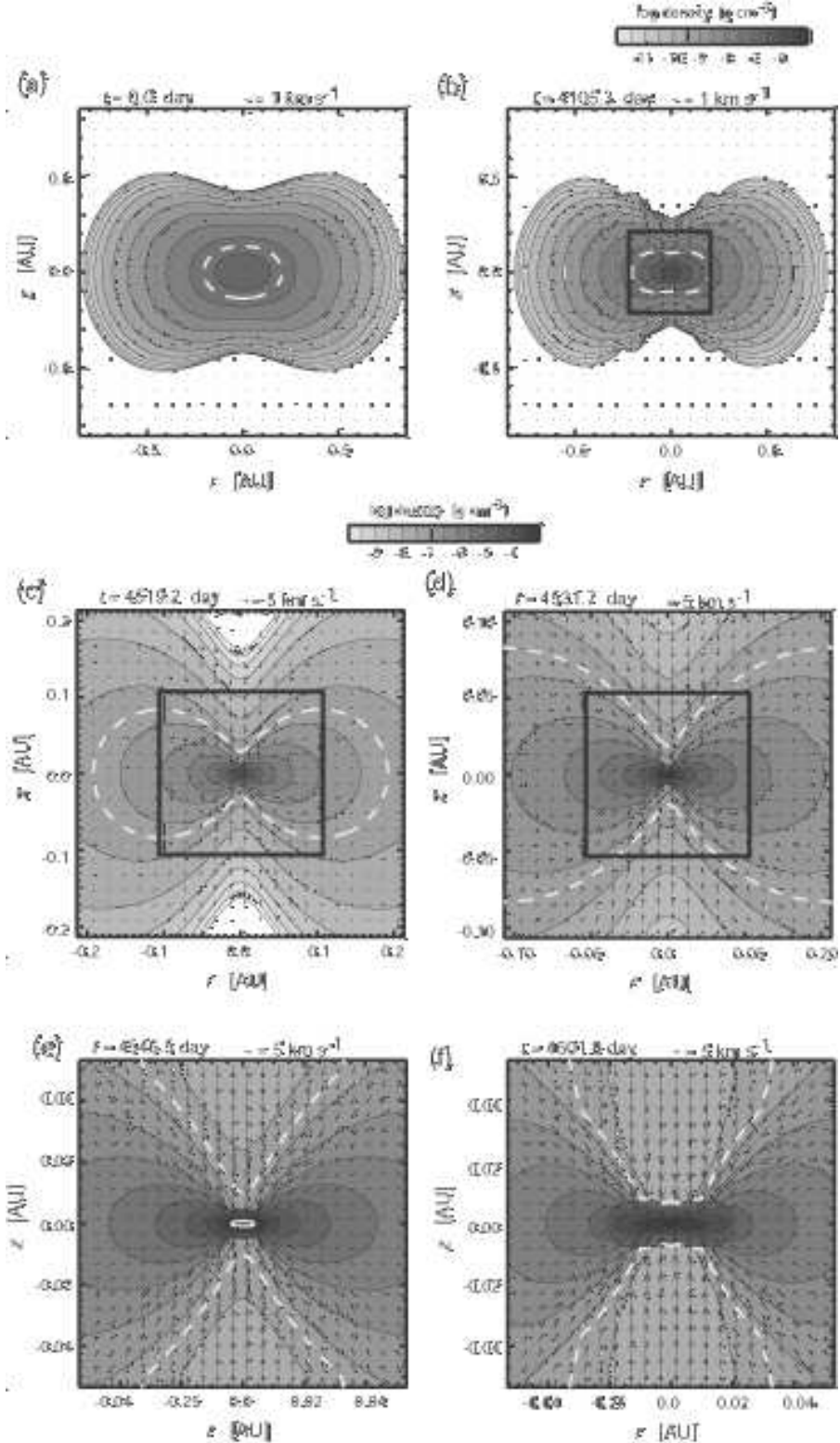}
\caption{Density and velocity distributions on the $r$-$z$ plane for Model A at six different stages.
The central density is ({\it a}) $\rho_{\rm c} \, = \, 2.3 \times 10^{-7}$ g  cm$^{-3}$, ({\it b}) $2.2 \times 10^{-6}$ g  cm$^{-3}$, ({\it c}) $2.9 \times 10^{-5}$ g  cm$^{-3}$, ({\it d}) $2.8 \times 10^{-4}$ g  cm$^{-3}$, ({\it e}) $1.81 \times 10^{-3}$ g  cm$^{-3}$, and ({\it f}) $1.0 \times 10^{-3}$ g  cm$^{-3}$ at each stage. 
The contour curves and grayness denote the density in the logarithmic scale. The arrows show the velocity distributions. 
\label{fig3}}
\end{figure}

\begin{figure}
\plotone{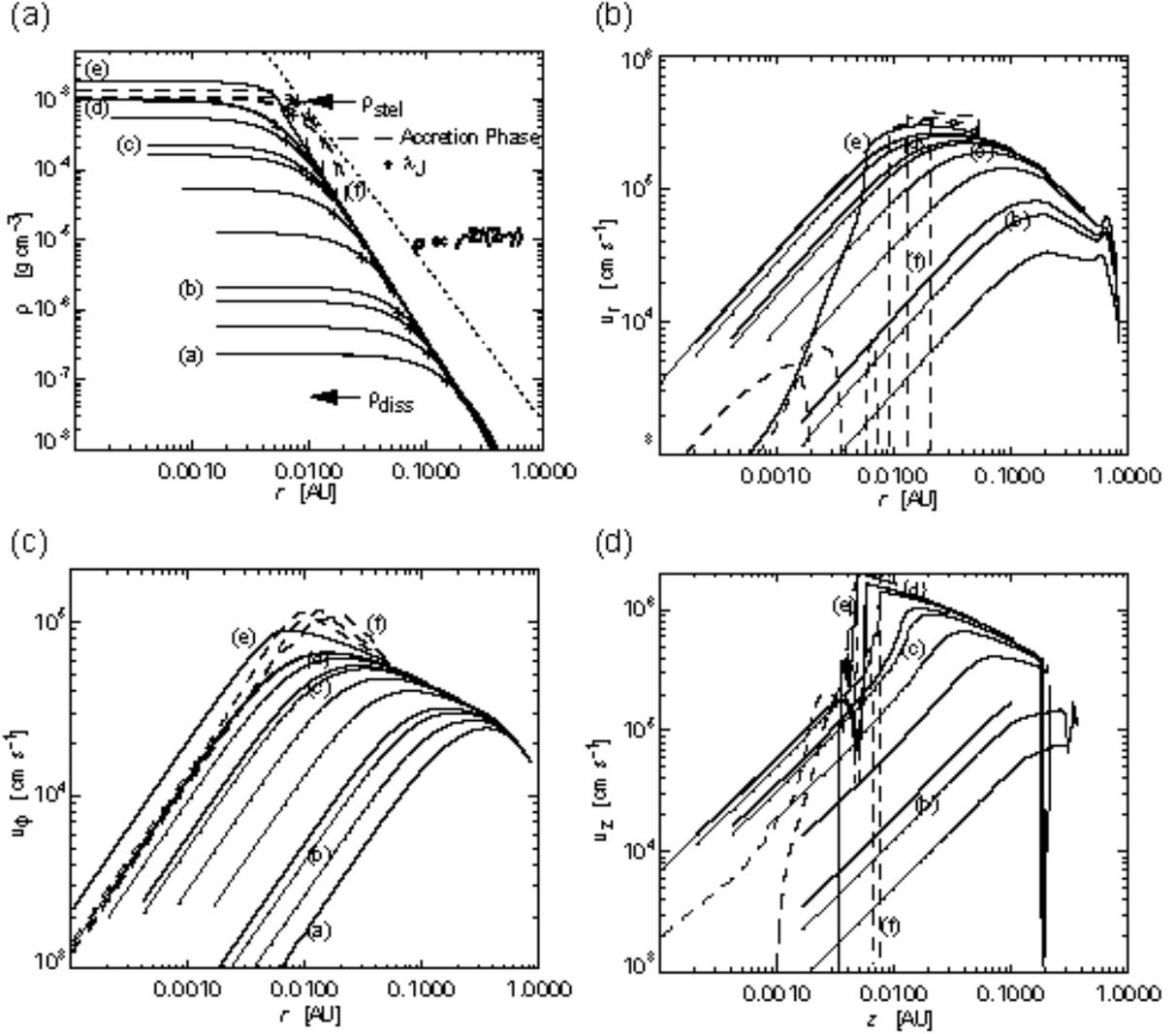}
\caption{({\it a})density, ({\it b}) radial infall velocity, and ({\it c})rotation velocity on the equatorial plane are shown for model A. Panel ({\it d}) shows the infall velocity along the $z$-axis. 
Solid curves represent the distribution in the second collapse phase. The evolution after the bounce is shown with dashed curves. 
Curves ({\it a})-({\it e}) correspond to the snapshots of Fig.3
\label{fig4}}
\end{figure}

\begin{figure}
\plotone{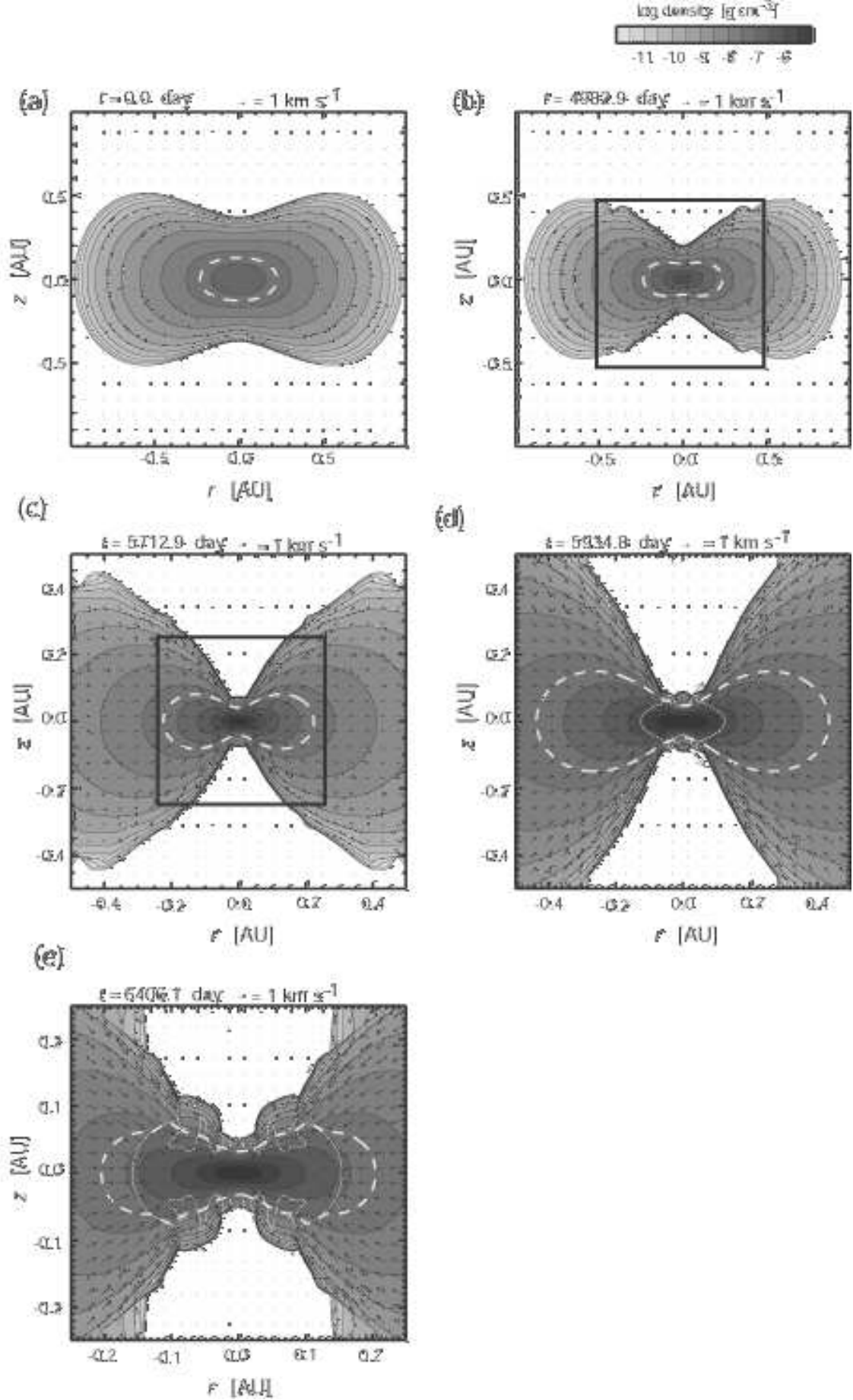}{f5.eps}
\caption{Same as Fig.3 but for model B. The white dotted curves denote the 1.5th core. \label{fig5}}
\end{figure}

\begin{figure}
\plotone{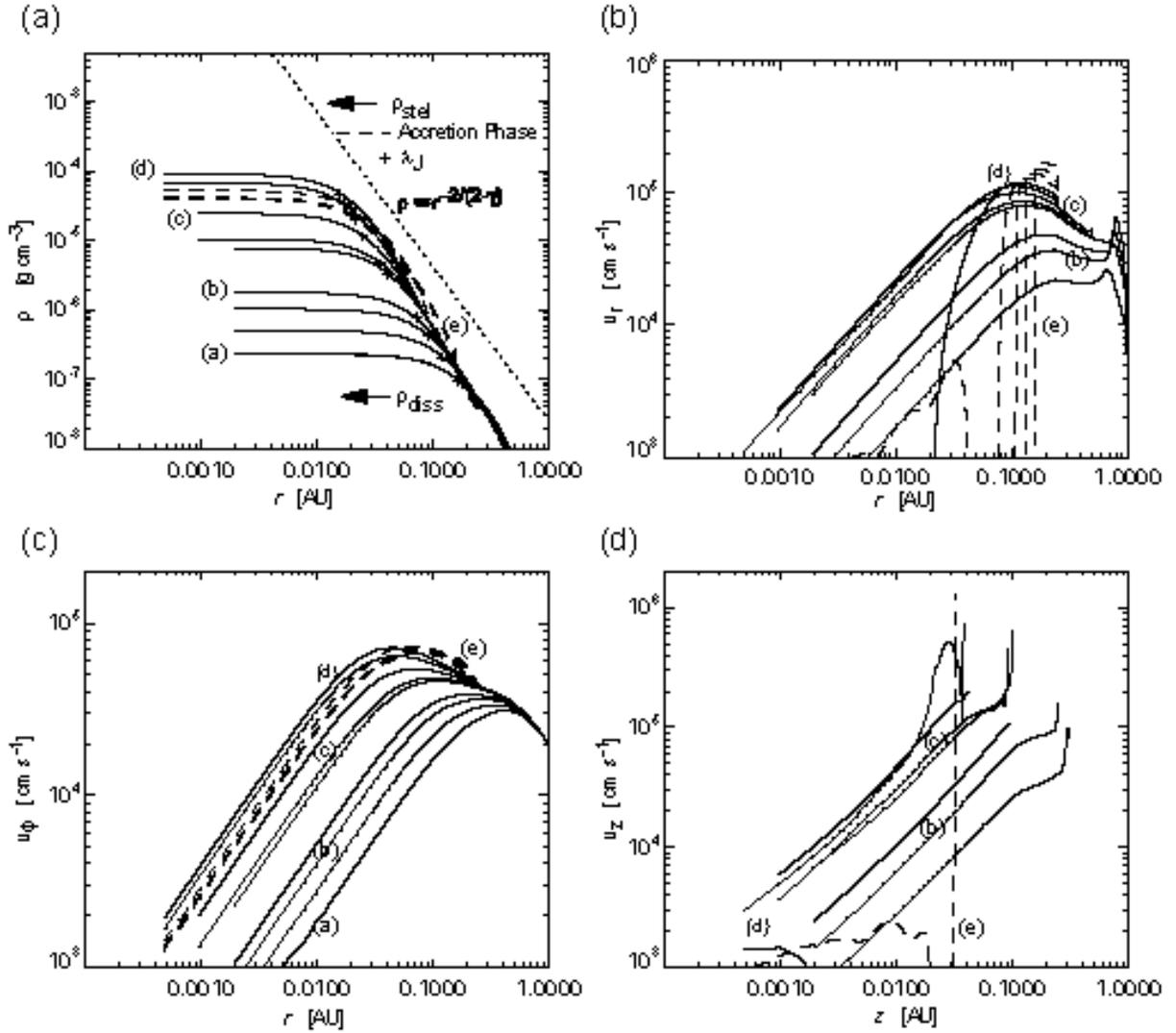}
\caption{Same as Fig. 4 but for model B. Curves ({\it a})-({\it a}) correspond to the snapshots of Fig. 5. \label{fig6}}
\end{figure}

\begin{figure}
\plotone{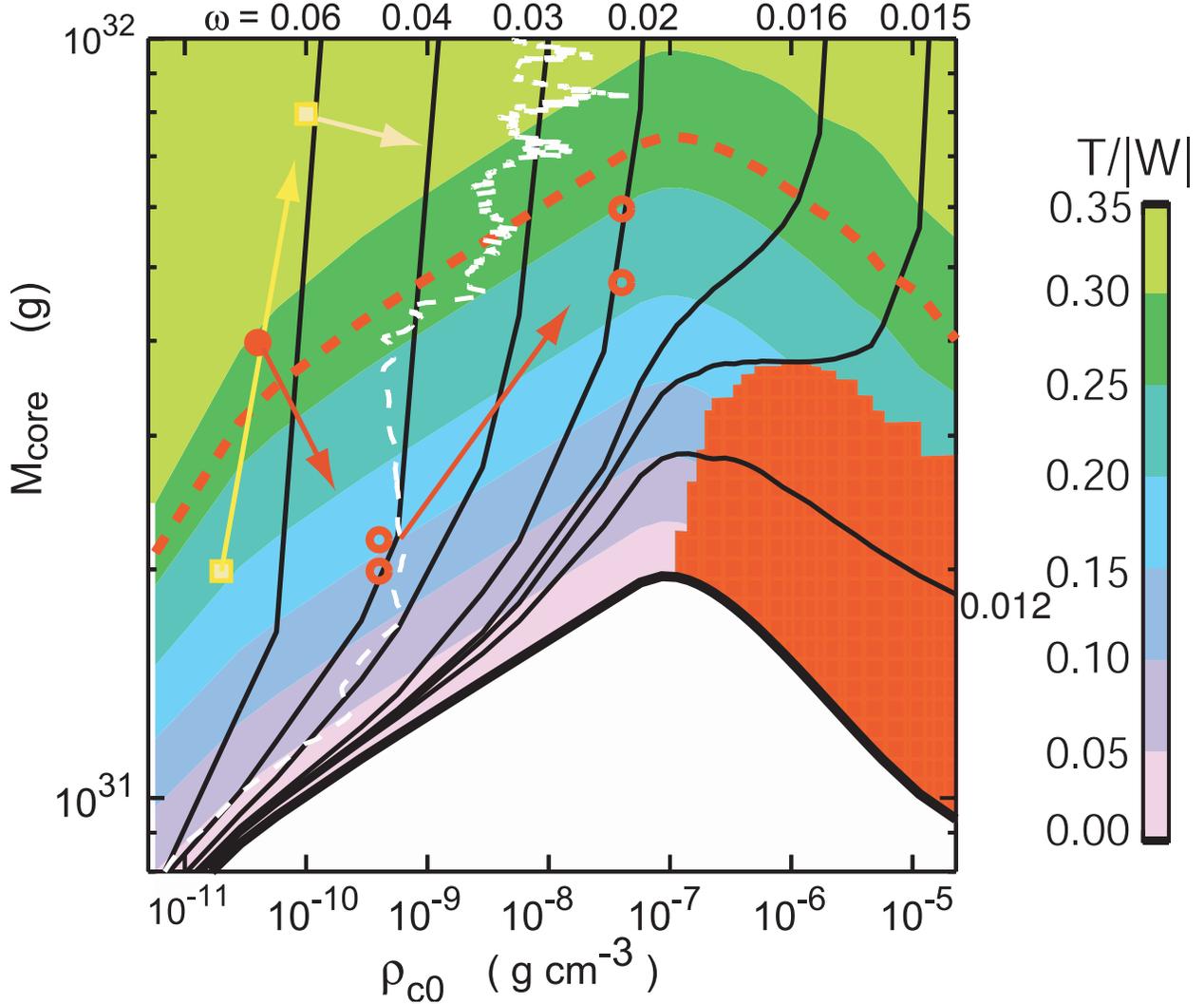}
\caption{Same as Fig. 2 but with rotational energy and three-dimensional hydro results. 
Color denotes the ratio of the rotational energy to the gravitational energy, $T/|W|$, and the red dashed curve denotes the critical value of $T/|W| \, = \, 0.273$ for the non-axisymmetric instability. 
The squares denote the results of Bate (1998).
The filled and open circles denote the results of HRM005 of MH03.
The first core for HRM005 model fragments into binary (open circles). 
The white dashed curve denotes the results for HRM003 of MH03. 
\label{fig7}}
\end{figure}

\begin{figure}
\plotone{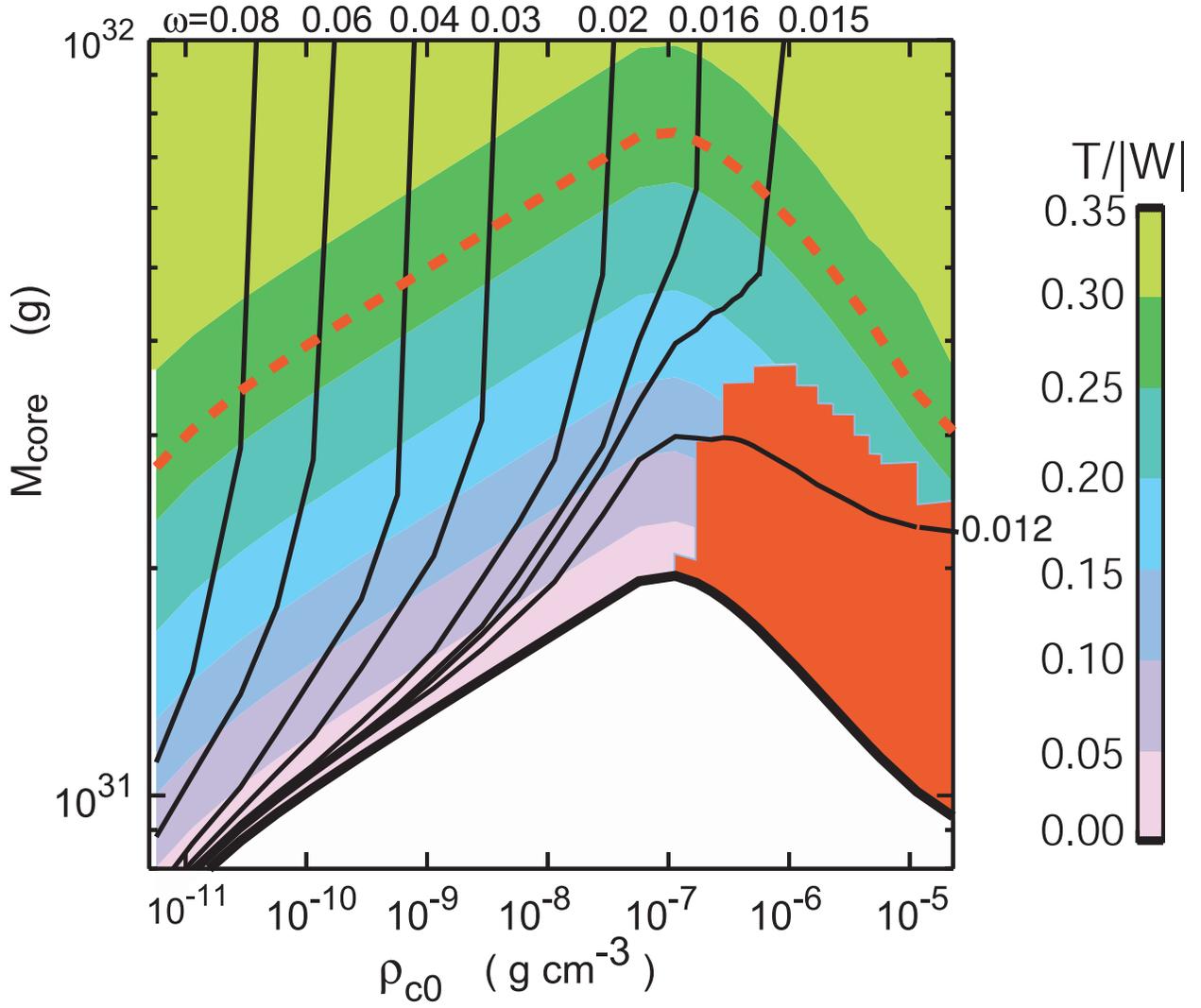}
\caption{Same as Fig.7 but for a different angular momentum distribution. The angular momentum distribution for the uniformly rotating, uniform density cylinder  is assumed. \label{fig8}}
\end{figure}

\begin{figure}
\plotone{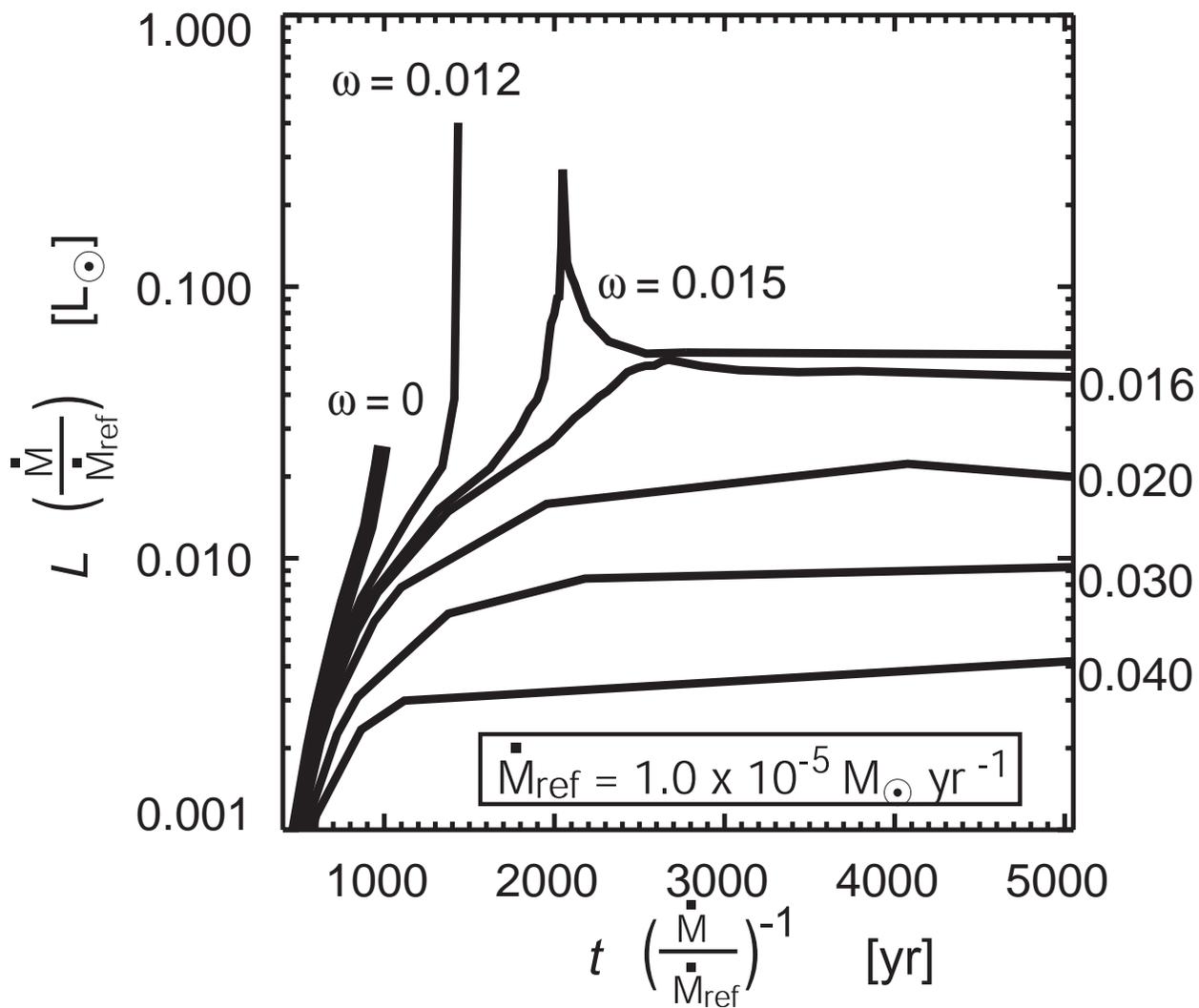}
\caption{Evolutionary curve of the first core luminosity during the first core phase. A constant mass accretion rate of $\dot{M} \, = \, 1.0 \times 10^{-5}$ M$_{\odot}$/yr is assumed as a reference. \label{fig9}}
\end{figure}

\end{document}